\documentclass[letter]{aa} 
\usepackage{graphicx}
\usepackage{txfonts}
\usepackage{rotating}
\usepackage{natbib}
\newcommand{\asec}{$^{\prime\prime}$}
\newcommand{\pas}{.\hskip-2pt$^{\prime\prime}$}

\def\i{I05345}
\def\H{N$_{2}$H$^{+}$}
\def\D{N$_{2}$D$^{+}$}

\def\kms{\mbox{km~s$^{-1}$}}
\def\cmc{cm$^{-3}$}
\def\cmq{cm$^{-2}$}

\def\solm{\mbox{M$_\odot$}}
\def\soll{\mbox{L$_\odot$}}
\def\Vlsr{$V_{\rm LSR}$}
\def\Dfrac{$D_{\rm frac}$}
\def\Tex{\mbox{$T_{\rm ex}$}}

\begin{document}

\title{Highly deuterated pre--stellar cores in a high-mass star formation region 
\thanks{Based on observations carried out with the IRAM Plateau de Bure Interferometer.
IRAM is supported by INSU/CNRS (France), MPG (Germany) and IGN (Spain).}}
\author{F. Fontani \inst{1} \and  P. Caselli \inst{2,3} \and
	T. L. Bourke \inst{4} 
	\and R. Cesaroni \inst{2}
	\and J. Brand \inst {1}
        }
\offprints{F. Fontani, \email{ffontani@ira.inaf.it}}
\institute{INAF-Istituto di Radioastronomia, Via Gobetti 101, 
           I-40129 Bologna, Italy \and
	   INAF-Osservatorio Astrofisico di Arcetri, Largo E. Fermi 5,
           I-50125 Firenze, Italy 
	   \and
	   School of Physics and Astrophysics, University of Leeds, Leeds, LS2 9JT, UK
	   \and
	   Harvard-Smithsonian Center for Astrophysics, 60 Garden Street MS42, Cambridge, MA 02138, USA
	   } 
\date{Received date; accepted date}

\titlerunning{Pre--stellar cores in I05345}
\authorrunning{Fontani et al.}

\abstract{}
{We have observed the deuterated gas in the high-mass star formation 
region IRAS 05345+3157 at high-angular resolution, in order to determine
the morphology and the nature of such gas.}
{We have mapped the \H\ (1--0) line with the Plateau de Bure Interferometer, 
and the \D\ (3--2) and \H\ (3--2) lines with the Submillimeter Array.}
{We have detected two condensations in \D , with masses
of $\sim 2-3$ and $\sim 9$ \solm\ and diameters of 0.05 and 0.09 pc, 
respectively. The high deuterium fractionation ($\sim 0.1$) and
the line parameters of the \D\ condensations indicate 
that they are likely low- to intermediate-mass 
pre--stellar cores, even though other scenarios are possible.}
{}
\keywords{Stars: formation -- Radio lines: ISM  -- ISM: individual (IRAS 05345+3157) -- ISM: molecules}

\maketitle
%

\section{Introduction}
\label{Introduction}

The initial conditions of the star formation process are still
matter of debate. Studies have begun to unveil the chemical and physical 
properties of starless low-mass cores on the verge of forming 
low-mass stars (Kuiper et al.~\citeyear{kuiper}; Caselli 
et al.~\citeyear{casellia};~\citeyear{casellib}; 
Tafalla et al.~\citeyear{tafalla02};~\citeyear{tafalla06}),
demonstrating that in the dense and cold nuclei of these cores
C-bearing molecular species such as CO and CS are strongly depleted
(e.g.~Caselli et al.~\citeyear{casellib}; Tafalla et al.~\citeyear{tafalla02}),
while N-bearing molecular ions such as \H\ and \D\
(Caselli et al.~2002a; 2002b; Crapsi et al.~2005) maintain
large abundances in the gas phase, and their column 
density ratio reaches values of $\sim 0.1$ or more, much higher than 
the cosmic [D/H] elemental abundance ($\sim 10^{-5}$, Oliveira et al.~2003).
 
The characterisation
of the earliest stages of the formation process of high-mass stars is more 
difficult than for low-mass objects, given their shorter evolutionary 
timescales, larger distances, and strong interaction with their 
environments. 
In order to check if the chemical properties peculiar of the
earliest stages of low-mass stars are valid also for high-mass stars,
Fontani et al.~(2006) observed the deuterated gas in 10 sources 
selected from two samples of high-mass protostar candidates
(Molinari et al.~1996; Sridharan et al.~2002), that 
are believed to be the closest to the earliest stages of the high-mass 
star formation process. 
One of these, IRAS 05345+3157 (hereafter \i ), stands out because 
of its very interesting characteristics: it is a luminous  
($1.38\times 10^{3}$\soll\ at a distance of 1.8 kpc, Zhang et al.~2005) 
young stellar object, embedded in a massive
($\sim 180$ \solm) dusty clump (Fontani et al.~2006)
in which Molinari et al.~(2002) revealed a complex structure in 
the molecular gas observed in the HCO$^{+}$(1--0) line. 
From IRAM-30m spectra of 
the \H\ (1--0) and \D\ (2--1) lines, Fontani et al.~(2006) have 
derived an average deuterium fractionation of $\sim 0.01$, 
quite close to the values found by Crapsi et al.~(2005) in 
low-mass starless cores. 
These results indicate the presence in the source
of molecular gas with physical conditions similar
to those of low-mass starless cores
(i.e. $T\sim 10$ K and $n_{\rm H_{2}}\sim 10^{6}$\cmc\ ).
However, the precise location and the distribution
of this gas, required to understand its nature, can be determined 
only through higher angular resolution observations. 

In this letter, we present observations of \H\ and \D\ towards \i , 
obtained with the Submillimeter Array (SMA) 
and the Plateau de Bure Interferometer (PdBI),
and we report on the detection of two compact condensations of \D\ that 
have chemical features typical of low-mass pre--stellar core 
candidates.  
A full report and a more detailed analysis of the data obtained will 
be presented in a forthcoming paper.


\section{Observations and data reduction}
\label{obs}


Observations of \D\ (3--2) (at 231321 MHz) and \H\ (3--2) (at 279511.7 MHz)
towards \i\ were carried out with the SMA\footnote{The Submillimeter Array 
is a joint project between
the Smithsonian Astrophysical Observatory and the Academia Sinica Institute
of Astronomy and Astrophysics, and is funded by the Smithsonian Institution 
and the Academia Sinica.} (Ho et al.~\citeyear{ho}) in the compact 
configuration on 30 January and 21 February 2007, respectively. 
The correlator was configured to observe
simultaneously the continuum emission and several other molecular
lines. The phase center was the nominal position of the
sub-mm peak detected with the JCMT (Fontani et al.~2006), 
namely R.A.(J2000)=05$^h$37$^m$52.4$^s$
and Dec.(J2000)=32$^{\circ}$00$^{\prime}$06\asec , and the local
standard of rest velocity \Vlsr\ is $-18.4$ \kms .
For gain calibration, observations of \i\ were alternated
with the sources 3C111 and J0530+135. 3C279 and Callisto were used 
for passband and flux calibration, respectively.
The SMA data were calibrated with the MIR package (Qi~2005),
and imaged with MIRIAD (Sault et al.~\citeyear{sault}). 
Channel maps were created with natural weighting, attaining a 
resolution of: 3\pas7$\times$3\pas0 for the \D\ (3--2) 
channel map; 3\pas0$\times$2\pas8 for the 225 GHz continuum image;
2\pas7$\times$2\pas0 for the \H\ (3--2) channel map;
1\pas9$\times$1\pas2 for the 284 GHz continuum image.


We observed the \H\ (1--0) line at 93173.7725 MHz towards \i\ with the 
PdBI on 11 and 20 August 2006, in the D 
configuration, and on 3 April 2007 in the C configuration.
We used the same phase reference and \Vlsr\ velocities as for
the SMA observations. The nearby point sources 0507+179 and 0552+398 were 
used as phase calibrators, while bandpass and flux scale were 
calibrated from observations of 3C345 and MWC349,
respectively. For continuum measurements, we placed two 
320 MHz correlator units in the band when making the observations in
D configuration, and six 320 MHz correlator units in C configuration. 
The \H\ lines were excluded in averaging these units to produce the 
final continuum image (at $\sim 96095$ MHz). The synthesized beam size of
the \H\ channel map was 3\pas2$\times$3\pas4, while that of the 
continuum was 3\pas1$\times$3\pas2. 
We stress that the observations of the \H\ (1--0) and \D\ (3--2) lines have
approximately the same angular resolution.
The data have been reduced with the GILDAS software,
developed at IRAM and the Observatoire de Grenoble.

%
The spectra of both \H\ and \D\ obtained from the cleaned
maps have been analysed with the software CLASS, with the method 
described in Sect.~2.1.1 of Fontani et al.~(2006).

\section{Results}
\label{res}

\subsection{Structure of the source}
\label{structure}

\begin{figure*}
\centerline{\includegraphics[angle=-90,width=14cm]{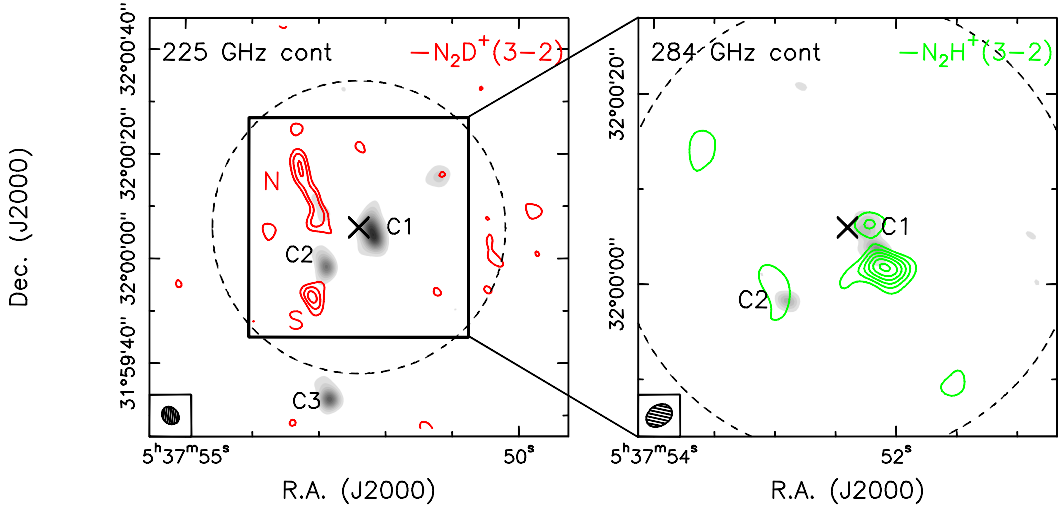}}
\caption{Left panel: map of the emission of the \D\ (3--2) line
integrated between --18.37 and --16.7 \kms , obtained with
the SMA towards \i\ (red contours). The two main condensations
are indicated as N and S. Contour levels start from the 
3$\sigma$ rms ($\sim 0.09$ Jy beam$^{-1}$), and are in steps of 2 $\sigma$.
The grey scale represents the 225 GHz continuum (first level is the 
3$\sigma$ rms = 0.006 Jy beam$^{-1}$; step = 3$\sigma$ rms). The three
compact continuum cores are indicated as C1, C2 and C3. The cross corresponds to 
the map center (\i\ nominal position) and the dashed circle represents the 
SMA primary beam at 225 GHz ($\sim 56$\asec ). 
The ellipse in the bottom left corner shows the synthesised beam of 
the \D\ image. 
\newline
Right panel: map of the \H\ (3--2) line emission integrated between
--19.6 and --16.1 \kms\ towards \i\ (green contours), observed with 
the SMA (first contour = 3$\sigma$ rms ($\sim 0.45$
Jy beam$^{-1}$); step = 2$\sigma$ rms).
The grey scale represents the 284 GHz continuum 
(first level is the 3$\sigma$ rms = 0.018 Jy beam $^{-1}$; step = 3$\sigma$ rms). 
The dashed circle represents the primary beam
at 284 GHz ($\sim 44$\asec ). 
The cross indicates the map center and the ellipse in
the bottom left corner is the SMA synthesised beam of the \H\ image.}
\label{n2dp32}
\end{figure*}

The map of the integrated intensity of the \D\ (3--2) line, superimposed
on the 225 GHz continuum map, is shown 
in the left panel of Fig.~\ref{n2dp32}. The map indicates that the
\D\ emission arises from two molecular condensations: an
extended clump located $\sim 10$\asec\ N--E of the map center and
elongated $\sim 15$\asec\ in the N--S direction, 
and a compact core, $\sim 5$\asec\ in size, located $\sim 15$\asec\ 
S--E of the map center. 
In the following we will identify these condensations as N 
and S, respectively, and they represent the targets of the 
present study. 
The 225 GHz continuum image shows two main compact cores inside the
SMA primary beam, one approximately corresponding to the map center
and the other located $\sim 10$\asec\ S--E of the map center.
Another compact source outside the interferometer primary beam is
located $\sim 35$\asec\ south of the map center, and it corresponds 
to the southern continuum source detected in the JCMT image by
Fontani et al.~(2006, see their Fig.~A.3). 
We will call these cores C1, C2 and C3, respectively.
None of these cores overlaps with the two \D\ condensations. Some extended
emission however is detected at $\sim 3\sigma$ towards N. The
right panel of Fig.~\ref{n2dp32} shows the \H\ (3--2) integrated emission 
superimposed on the 284 GHz continuum emission, both observed with
the SMA. At this frequency the continuum source C1 is resolved into a 
main peak and a fainter secondary peak north of the main one. 
The \H\ (3--2) line integrated emission is compact and
well overlaps with core C1, while it is detected only at 3$\sigma$ 
towards C2, and undetected towards N and S.
The detection of \H\ towards C1, and the non-detection of \D\ at 
that location, is probably due to the fact that the continuum image is
dominated by the emission of the warm dust, so that in the 
continuum condensation the temperature is too high for
significant deuterium fractionation. 

\begin{figure*}
\centerline{\includegraphics[angle=-90,width=12.5cm]{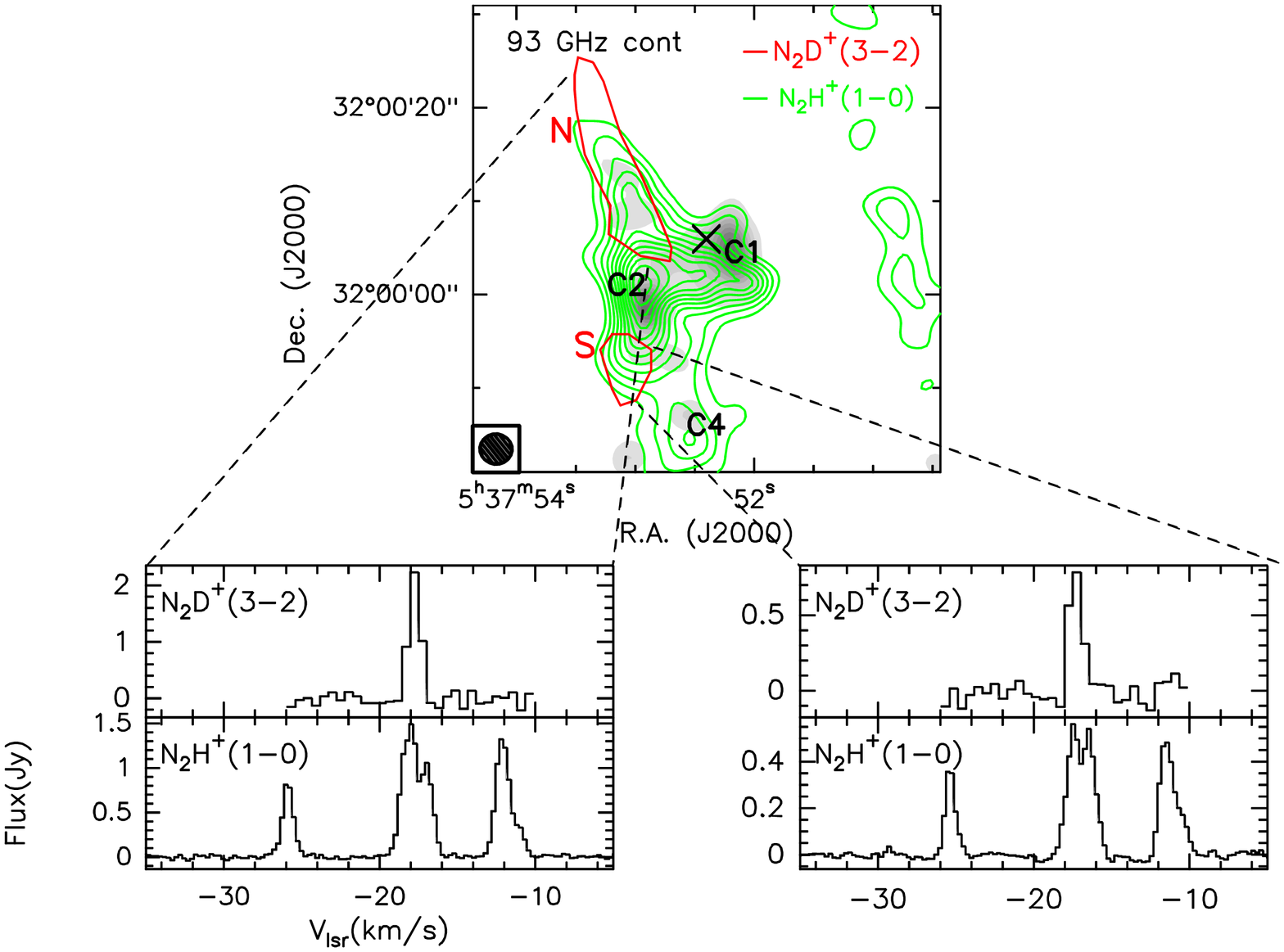}}
\caption{Map of the intensity of the \H\ (1--0) line
(green contours) integrated between --21.2 and --14.5 \kms , corresponding
to the main group of the hyperfine components, 
observed with the PdBI. Levels range from
the 3$\sigma$ rms, which is $\sim 0.02$ Jy beam$^{-1}$, to 0.26 
Jy beam$^{-1}$, in steps of 3$\sigma$. 
The grey scale represents the 97 GHz continuum: the levels range from
the 3$\sigma$ rms (4.2$\times 10^{-4}$ Jy beam$^{-1}$) to 4$\times 10^{-3}$ 
Jy beam$^{-1}$, in steps of 3$\sigma$. Below the map,
the spectra of \D\ (3--2) and \H\ (1--0) integrated over the red contours, 
which correspond to the 3$\sigma$ level of the \D\
condensations N and S (see Fig.~\ref{n2dp32}),
are shown.
The other symbols have the same meaning as in left panel of 
Fig.~\ref{n2dp32}. The continuum source identified
in Fig.~\ref{n2dp32} as C3 is out of the box.} 
\label{n2hp10}
\end{figure*}

\begin{table*}
\tiny
\caption{\H\ (1--0) and \D\ (3--2) line parameters. Between parentheses, 
the uncertainties of the fitting procedure (see text) are given.}
\label{tab_lin}
\begin{center}
\begin{tabular}{cccccccc}
\hline \hline
     & Spectral resolution & velocity range  & $\int T_{\rm MB}{\rm d}$V & $V_{\rm LSR}$ & FWHM & $\tau_{\rm m}$ & \Tex\  \\
       &(\kms ) & (\kms ) & (K \kms ) & (\kms ) & (\kms ) & & (K)  \\
\hline
 N     &  & &  &  &  & & \\
 \H\ (1--0) & 0.25  & --29.9; --12.5 & 8.7(0.3) &  -17.915(0.003)  &     0.732(0.007) &   4.132(0.212)  &    6.7(0.1) \\
 \D\ (3--2) & 0.5 & --18.3; --16.7 & 0.64(0.03)  &  -17.91(0.02) &     0.88(0.014) &  0.1(fixed)   &   6.7$^{1}$  \\
 S     &  & &  &  &  & & \\
\H\ (1--0) & 0.25 & --26.5; --9.3 & 10.3(0.5) &   -17.283(0.005)  &    0.73(0.01) &   6.1(0.4)  &   6.6(0.1) \\
\D\ (3--2) & 0.5 & --18.3; --16.7 & 0.82(0.06) &  -17.51(0.04)  &     0.88(0.06) &     0.1(fixed)  &  6.6$^{1}$ \\
\hline
\end{tabular}
\end{center}
$^{1}$ assumed to compute the \D\ total column density in Table~\ref{tab_dfrac}
\end{table*}
\normalsize
The integrated intensity map of the main group of
the hyperfine (hfs) components of the \H\ (1--0) line (see e.g. Caselli et
al.~\citeyear{caselli95}) observed with the PdBI is shown in 
Figure~\ref{n2hp10}. The emission
is extended and irregular, covering an area of $\sim 40$\asec\
in the N-S direction, and $\sim 20$\asec\ in the W-E direction.
The main emission peak is close to C2, and the shape of the emission 
indicates that \H\ (1--0) and the 93 GHz continuum are fairly-well 
overlapping. An extended component in both line and continuum
oriented NE-SW roughly corresponds to
condensation N. Another compact source, C4, is detected $\sim 20$\asec\ 
south of the map center.
Such a different distribution of the integrated intensity 
with respect to that of the \H\ (3--2) line is
probably due to the fact that the extended emission is
filtered out differently by the two interferometers. To compute the
amount of the missing flux, we have compared our interferometric
spectra with the single-dish spectra obtained with the IRAM-30m telescope
(see Fontani et al.~2006). We have resampled the single-dish 
spectra of \H\ (1--0) and (3--2) to the same resolution in velocity 
as the interferometric spectra. 
In the \H\ (1--0) line,
the flux measured by the PdBI is $\sim 2$ times less than that
measured with the IRAM-30m telescope, while in the
\H\ (3--2) line the flux measured with the SMA is only 
one fifth of that measured with the 30m antenna. This indicates
that the extended emission is much more resolved out in
the SMA map with respect to the PdBI map.

\subsection{\H\ and \D\ line parameters and deuterium fractionation}
\label{dfrac}

In Table~\ref{tab_lin} we give the \H\ (1--0) and
\D\ (3--2) line parameters of the spectra integrated over the 3$\sigma$ rms
level of the \D\ emission in condensations N and S: 
in \mbox{Cols.~3 -- 7} we list integrated 
intensity ($\int T_{\rm MB}{\rm d}$V), peak velocity ($V_{\rm LSR}$), FWHM,
opacity of the main component ($\tau_{\rm m}$), and excitation temperature
(\Tex ) of the lines. The integrated intensities have been
 computed over the velocity range given in Col.~2, while 
for the other parameters we have adopted the fitting 
procedure described in Sect.~2.1.1 of Fontani et al.~(2006). 
We do not give the line optical depths
and \Tex\ for the \D\ lines because in both spectra the uncertainties 
are comparable to the values obtained. Because of this, we have fitted
the lines forcing the optical depth to be 0.1.

Linewidths and peak velocities of both
\H\ and \D\ are very similar in both condensations, indicating that
they are tracing the same gas. In particular,
the lines are $\sim 0.8$ \kms\ broad for both molecular
species, i.e. nearly half of that
observed with the IRAM-30m telescope. This can be due either
to extended components with different velocity that have been resolved out,
or to a decrease in turbulence going from the pc-scale to the 
sub-pc scale. However, it is interesting to
notice that the lines are broader than the typical \H\ and \D\ 
lines observed towards low-mass pre--stellar cores, 
for which values of $\sim 0.2 - 0.3$ \kms\ are found on 
comparable linear scales (Crapsi et al. 2005, Roberts \& Millar 2007).

We have derived \H\ and \D\ total column densities, $N$(\H ) and
$N$(\D ), following the method described in the Appendix 
of Caselli et al.~(\citeyear{casellib}), which assumes a constant
\Tex . $N$(\H ), $N$(\D ) and \Dfrac = $N$(\D )/$N$(\H ),
derived for N and S, are listed in Cols. 2, 3 and 4 of Table~\ref{tab_dfrac},
respectively. These have been obtained using the values of \Tex\
given in Table~\ref{tab_lin}. For the \D\ (3--2) lines,
we have assumed the excitation temperature
of the \H\ (1--0) line because we could not derive a reliable
value from the fitting procedure taking into account the
line hyperfine splitting, as
already pointed out. In both condensations, \Dfrac\
is $0.11$, which is comparable to 
the values of \Dfrac\ found in low-mass pre--stellar cores 
by Crapsi et al.~(2005), following the same method. 
These values are also comparable to those derived by
Pillai et al. (2007) in infrared-dark clouds from deuterated 
ammonia. However, their observations are related to the very
cold, pc-scale molecular envelope, and not to compact sub-pc 
scale cores. 

\subsection{Nature of the \D\ condensations}
\label{discu}

The main finding of this work is that the \D\ emission in \i\
is concentrated in two condensations, both of them characterised
by high values of deuterium fractionation.
We discuss now which is the nature of these condensations.

We have computed the angular diameters of both condensations 
assuming that the \D\ integrated intensity profile can be fitted
with a 2D Gaussian: the geometric mean of the 
major and minor axis resulting from these fits has
been corrected for the beam size. Then, the linear diameters, $L$,
have been computed using a source distance of 
1.8 kpc (Zhang et al.~2005).
Then, the mass of an equivalent homogeneous sphere of 
diameter $L$, $M_{\rm N_2H^+}$, has been obtained from $N$(\H )
assuming a \H\ average abundance of 1.5$\times 10^{-10}$ 
(Fontani et al.~2006). $L$ and $M_{\rm N_2H^+}$ are listed
in Cols. 5 and 6 of Table~\ref{tab_dfrac}, respectively.
For N and S, we derive $M_{\rm N_2H^+}\sim 8.7$ and $2.5 M_{\odot}$,
and $L\sim 0.09$ and 0.05 pc, respectively. 
These values, together with a \Dfrac\ $\simeq 0.1$, 
suggest that both cores may be low-mass pre--stellar cores.
On the other hand, as stated in Sect.~\ref{dfrac}, both N and S show
lines broader than those typically observed in low-mass
pre--stellar cores. 
We propose three possible scenarios about the nature of the two \D\
condensations: (i) they represent the residual of the 
extended cold cloud in which the cluster of young stellar
objects was formed, whose physical/chemical 
conditions have not been altered yet by the cluster members;
(ii) they are a group of unresolved low-mass starless cores;
(iii) they are single starless cores, which are either 
low- to intermediate-mass pre--stellar cores, or the seeds of 
future massive forming stars.

The first scenario is very unlikely for both N and S, 
since the \D\ (3--2) line is
expected to trace gas denser than $\sim 10^6$ \cmc , while the 
H$_2$ volume densities measured in the
pc-scale gas associated with high-mass star formation regions
typically reach values no higher than $\sim 10^4 - 10^5$ \cmc .
The third scenario is the most probable for S, while for N,
which shows an elongated structure (see Fig.~\ref{n2dp32}) 
indicative of possible multiple low-mass components, the
second scenario is also possible. In fact, the shape of N
resembles that of Oph A, a molecular clump with elongated
shape, resolved into several low-mass starless cores (Andr\'e et al.
2007). If so, the observed broad lines could be simply due to unresolved 
cores with diffent velocity. A detailed comparison between N and 
Oph A will be done in the forthcoming paper. 
However, it is also possible that N is, like S, a 
unique condensation, the observed structure being due to 
shaping by a powerful outflow associated with another cluster member.
In fact, an outflow driven by the massive source C1 ($\sim 20$ \solm )
has been detected observing the CO (2--1) line
(this line is part of the
large dataset obtained with the SMA that will be published in the 
forthcoming paper),
and the red lobe is indeed detected at the edge of condensation N. 
If both N and S are single cores, what may cause the
observed broad lines? 
The larger turbulence could be simply understood as the
result of a high pressure environment (see e.g. McKee \& 
Tan~\citeyear{mckee}). 
Despite of this, it is not at all clear which fraction of the line width
is actually broadened by systematic motions, in particular infall or
accretion of molecular material onto the pre-stellar cores. Only high
resolution kinematic studies of the large scale gas around these
condensations will give us clues on the line width partition, thus providing
important constraints on the dynamical evolution of massive star forming
regions.
\begin{table}
\caption{\H\ and \D\ total column densities ($N$(\H )
and $N$(\D )), deuterium fractionation (\Dfrac ), linear
diameter ($L$), and mass derived from $N$(\H ) ($M_{\rm N_2H^+}$)
for the \D\ condensations N and S of Figure~\ref{n2dp32}. 
The uncertainties computed following
the standard propagation of the errors are given between parentheses.}
\label{tab_dfrac}
\begin{center}
\begin{tabular}{cccccc}
\hline \hline
source & $N({\rm N_2H^+})$ & $N({\rm N_2D^+})$ & $D_{\rm frac}$ & $L$ & $M_{\rm N_2H^+}$ \\
       & ($\times 10^{13}$\cmq )  & ($\times 10^{12}$\cmq )  & & (pc) & ($M_{\odot}$) \\
\hline
N    &  1.9(0.6) &  2.1(0.3) &    0.11(0.04) & 0.09 & 8.7 \\
S  &   1.5(0.5) &  1.6(0.3)  &    0.11(0.04) & 0.05 & 2.5 \\
\hline
\end{tabular}
\end{center}
\end{table}
\newline
\newline
It is a pleasure to thank the staff of the Smithsonian Astrophysical
Observatory for the SMA observations. We also thank
the IRAM staff for his help in the calibration of the PdBI data.
Many thanks to the anonymous Referee for his/her useful comments
and suggestions.
{}


\begin{thebibliography}{}

\bibitem[Andr\'e et al.(2007)]{andre}
Andr\'e, Ph., Belloche, A., Motte, F., Peretto, N, 2007, A\&A, 472, 519 
\bibitem[Beuther et al.(2007)]{beuther}
Beuther, H., Churchwell, E.B., McKee, C.F., Tan, J.C. 2007, Protostars and Planets V, p. 165
\bibitem[Caselli et al.(1995)]{caselli95}
Caselli, P., Myers, P.C., Thaddeus, P. 1995, ApJ, 455L, 77
ApJ, 523, L165
\bibitem[Caselli et al.(2002a)]{casellia}
Caselli, P., Walmsley C.M., Zucconi, A. et al. 2002a, ApJ, 565, 331
\bibitem[Caselli et al.(2002b)]{casellib}
Caselli, P., Walmsley C.M., Zucconi, A. et al. 2002b, ApJ, 565, 344
\bibitem[2005]{crapsi}
Crapsi, A., Caselli, P., Walmsley, C.M., et al. 2005, ApJ, 619, 379
\bibitem[2006]{fonta06}
Fontani, F., Caselli, P., Crapsi, A., et al. 2006, A\&A, 460, 709
\bibitem[Ho et al.(2004)]{ho}
Ho, P.T.P., Moran, J.M. \& Lo K.Y. 2004, ApJ, 616, L1
\bibitem[Kuiper et al.(1996)]{kuiper}
Kuiper, T.B.H., Langer, W.D., \& Velusamy, T. 1996, ApJ, 468, 761
\bibitem[McKee \& Tan(2002)]{mckee}
McKee, C.F. \& Tan J.C. 2002, Nature, 416, 59
\bibitem[1996]{mol96}
Molinari, S., Brand, J., Cesaroni, R., Palla, F. 1996, A\&A, 308, 573
\bibitem[2003]{oliveira}
Oliveira, C. M., H\'ebrard, G., Howk, J. C., et al. 2003, ApJ, 587, 235
\bibitem[2007]{pillai}
\mbox{Pillai, T., Wyrowski, F., Hatchell, J., Gibb, A.G., Thompson, M.A. 2007, 467, 207}
\bibitem[2005]{qi}
\mbox{Qi, C. 2005, The MIR Cookbook (http://cfa-www.harvard.edu/~cqi/mircook.html)}
\bibitem[2007]{rem}
Roberts, H. \& Millar, T.J.007, A\&A, 471, 849
\bibitem[Sault et al(1995)]{sault}
Sault, R.J., Teuben, P.J. \& Wright, M.C.H. 1995, ASPC, 77, 433
\bibitem[2002]{sridharan}
Sridharan, T.K., Beuther, H., Schilke, P., Menten, K.M., Wyrowski, F. 2002, ApJ, 566, 931
\bibitem[Tafalla et al.(2002)]{tafalla02}
Tafalla, M., Myers, P.C., Caselli, P., Walmsley, C.M., Comito, C. 2002, ApJ, 
569, 815
\bibitem[Tafalla et al.(2006)]{tafalla06} 
Tafalla, M., Santiago, J., Myers, et al.~2006, A\&A, 455, 577 
\bibitem[2005]{zhang}
Zhang, Q., Hunter, T.R., Brand, J. et al. 2005, ApJ, 625, 864

\end{thebibliography}
\end{document}